# AoECR: AI-ization of Elderly Care Robot

Linkun Zhou, *Jian Li*, Yadong Mo, Xiangyan Zhang, Ying Zhang and Shimin Wei

*Abstract*—Autonomous interaction is crucial for the effective use of elderly care robots. However, developing universal AI architectures is extremely challenging due to the diversity in robot configurations and a lack of dataset. We proposed a universal architecture for the AI-ization of elderly care robots, called AoECR. Specifically, based on a nursing bed, we developed a patient-nurse interaction dataset tailored for elderly care scenarios and fine-tuned a large language model to enable it to perform nursing manipulations. Additionally, the inference process included a self-check chain to ensure the security of control commands. An expert optimization process further enhanced the humanization and personalization of the interactive responses. The physical experiment demonstrated that the AoECR exhibited zero-shot generalization capabilities across diverse scenarios, understood patients' instructions, implemented secure control commands, and delivered humanized and personalized interactive responses. In general, our research provides a valuable dataset reference and AI-ization solutions for elderly care robots.

*Index Terms*—AI-ization, AoECR, autonomous interaction, elderly care, large language model.

## I. Introduction

Based on the most recent data from the United Nations, the global population is anticipated to escalate to 8.5 billion by the year 2030, with projections reaching 9.7 billion by 2050. In tandem, the demographic of individuals aged 65 and above is expected to surge to 2.2 billion by the 2070s[1]. Furthermore, the number of elderly individuals with physical frailty, who are unable to partake in physical activity, is predicted to exceed 440 million[2]. The phenomenon of global aging, coupled with a declining birth rate, is intensifying the demand for elderly care services. However, statistics from 2019 reveal that the number of nursing staff specialized in elderly care constituted merely 9% of the total professional nursing workforce, amounting to nearly 15 million individuals globally[3]. The demographic aged 65 and older is projected to increase significantly over the forthcoming three decades. This significant mismatch between the escalating demand for nursing care and the existing resources has emerged as a critical global challenge[4]. To address these challenges, the demand for elderly care robots is increasing[5]. This approach helps to bridge the gap between demand and resources and enhances the well-being of elderly individuals with compromised functional abilities.

The current state of research on elderly care robots indicates that these robots are playing an increasingly important role in assisting the daily lives of the elderly. They integrate advanced robotics technology to meet the needs of the elderly in terms of medical care, daily living activities, and emotional communication. Researchers have developed various types of elderly care robots, including intelligent robots based on medical functions and physiological index monitoring[6], robots for daily living assistance and rehabilitation training[7], and companion robots based on emotional interaction[8]. These robots can provide comprehensive services, from health monitoring and assistance with daily living tasks to emotional support[9].

Despite significant advancements in elderly care robots, there are still several challenges in research and application. Firstly, the safety of human-robot interaction is a critical issue, especially in terms of physical contact and psychological impact. Robots must be able to accurately perceive human requests and provide a comfortable interactive experience[10]. Secondly, the acceptance of robot technology by the elderly poses a challenge, as the complexity of operations and unfamiliarity with technology may lead to their resistance to using these technologies[11].

In this paper, we focus on the control of an elderly care nursing bed. Initially, we constructed a Patient-Nurse Interaction (PN-I) dataset. We generated data pairs using zero-shot learning. Additionally, considering the expression issues of the elderly, we expanded the dataset to incorporate samples that include logical disorientation, inarticulateness, and stuttering. Based on the PN-I dataset, we fine-tuned a main LLM using the LoRA[12]. With appropriate prompts, the main LLM can generate tentative control commands and responses. Furthermore, we designed a CoS (Chain of Self-check) to ensure the security and accuracy of control commands.

Manuscript received Month xx, xxxx; revised Month xx, xxxx; accepted Month x, xxxx. This work was supported by the National Key Research and Development Program of China（Nos.2023YFB4706100, 2023YFB4706102）, the interdisciplinary Team of Intelligent Elderly Care and Rehabilitation in the "Double first-class" Construction of Beijing University of Posts and Telecommunications (No. 530324004) and the Fundamental Research Funds for the Central Universities of BUPT (NO. 510224074).

Linkun Zhou is with the Laboratory of Robotics Mechanism and Cross Innovation, Beijing University of Posts and Telecommunications, Beijing, CO 102206 CHN (e-mail: 3215681611@bupt.edu.cn).
Jian Li is with the Laboratory of Robotics Mechanism and Cross Innovation, Beijing University of Posts and Telecommunications, Beijing, CO 102206 CHN (e-mail: jianli_628@126.com).
Yadong Mo is with the Laboratory of Robotics Mechanism and Cross Innovation, Beijing University of Posts and Telecommunications, Beijing, CO 102206 CHN (e-mail: yadong_mo@163.com).
Xiangyan Zhang is with the Laboratory of Robotics Mechanism and Cross Innovation, Beijing University of Posts and Telecommunications, Beijing, CO 102206 CHN (e-mail: zhangxiangyan1994@163.com).
Ying Zhang is with the Laboratory of Robotics Mechanism and Cross Innovation, Beijing University of Posts and Telecommunications, Beijing, CO 102206 CHN (e-mail: yingzhang_bupt@bupt.edu.cn).
Shimin Wei is with the Laboratory of Robotics Mechanism and Cross Innovation, Beijing University of Posts and Telecommunications, Beijing, CO 102206 CHN (e-mail: wsmly@bupt.edu.cn).

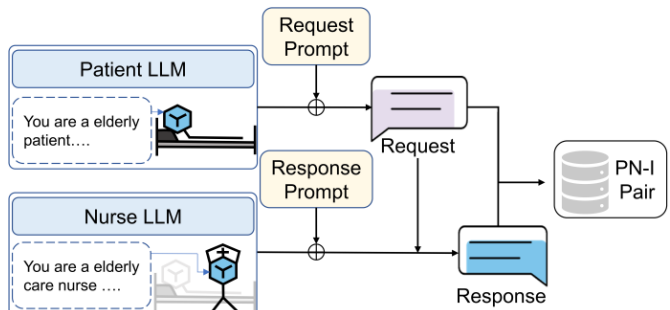

**Fig. 1.** Simulated dialogue environment and zero-Shot generation process

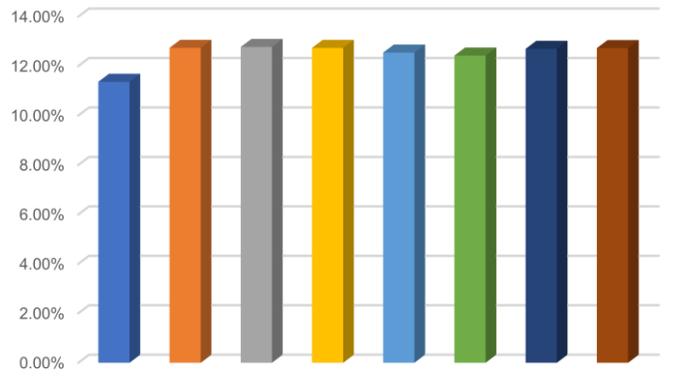

**Fig. 2.** Action type statistics in PN-I dataset

Meanwhile, tentative responses were optimized through an expert optimization to refine them more humanized and personalized. Ultimately, through a series of quantitative assessment experiments and physical experiments, we validated that the AoECR can accurately recognize patient requests, generate safe and correct control commands, and provide humanized and personalized responses.

In summary, our main contributions are as follows:
1. We presented a universal AoECR architecture that makes elderly care robot AI-ization.
2. We created a patient-nurse interaction dataset to fine-tune an LLM, which can serve as a reference for further research of elderly care.
3. We designed a CoS pipeline and an expert optimization pipeline to ensure the correctness and safety of control commands, as well as to enhance the humanization and personalization of interactive responses.
4. Experiments demonstrated the interactive ability of AoECR.

## II. RELATED WORK

### A. The Large Language Model

The development of LLMs originated from the concept of pre-trained language models[13], which are trained on large-scale unlabeled text to capture basic language structures and fine-tuned on task-specific datasets to adapt to different application scenarios. This paradigm of "pre-training and fine-tuning", such as GPT-2[14] and BERT[15], promotes the development of diverse and effective model architectures. With the advancement of technology, LLMs such as GPT-3[16], GPT-4[17], PaLM[18], and LLaMA[19] have been trained on large-scale text corpora with billions of parameters and aligned with human values after initial pre-training to understand better and execute human commands[20].

### B. Fine-tuning to Medical and Elderly Care.

Researchers have extensively studied fine-tuning technologies, such as LORA and QLORA[21], and adapters to allow fine-tuning with a few model parameters updated[22], thereby reducing computational costs[23]. The development of these technologies enables LLMs to adapt to specific fields[24], such as the medical field, by fine-tuning the injection of medical knowledge to enhance their capabilities in the medical field[25].

For instance, DoctorGLM is a Chinese consultation model based on ChatGLM-6B[26] that has been fine-tuned and deployed with a Chinese medical dialogue dataset[27], utilizing methods such as LoRA and P-Tuning v2[28]. Additionally, the BenTsao project has open-sourced a set of large language models fine-tuned with Chinese medical instructions, including LLaMA, Alpaca Chinese[29], and Bloom[30]. This initiative has improved the question-answering capabilities of the base model within the medical field.

## III. THE PATIENT-NURSE INTERACTIVE DATASET

Due to the shortage of clinical application cases, inconsistency in data quality, and the protection of patient privacy, we find out that the creation of real-world elderly care datasets is difficult. To overcome these difficulties, we have constructed a simulated environment for patient-nurse. By employing zero-shot learning, we prompted the LLMs to generate a series of patient requests and nurse responses. Furthermore, to make the dataset more suitable for elderly care, we extended the dataset based on the different clarity of expression. Ultimately, we successfully constructed a patient-nurse interaction Dataset, named PN-I.

### A. Simulated environment to zero-shot generation

Zero-shot is an innovative technology in the field of machine learning, aimed at constructing classifiers capable of recognizing categories not encountered during the training phase. In the context of large language models, zero-shot refers to the ability of models to perform specific tasks solely based on task prompts. In Fig. 1, we deployed two large language models within a simulated dialogue environment, designated as a patient LLM and a nurse LLM. Our prompts included detailed descriptions of the clinical environment, such as information about the nursing bed. Subsequently, we provided each LLM with different task prompts. The patient LLM, given by various reasons and specific control commands, should make requests to adjust the nursing bed, while the nurse LLM should make simple responses. In the simulated environment, each execution enabled the zero-shot generation of patient-nurse interaction data pairs. To enrich the generated data pairs, we provided the patient LLM with various reasons for adjusting the nursing bed, such as direct adjustment needs, physical discomfort, and psychological sensations, among others. Ultimately, we constructed an interaction dataset comprising nearly 40,000 data pairs. Statistical results, based on different actions, are shown in Fig. 2.

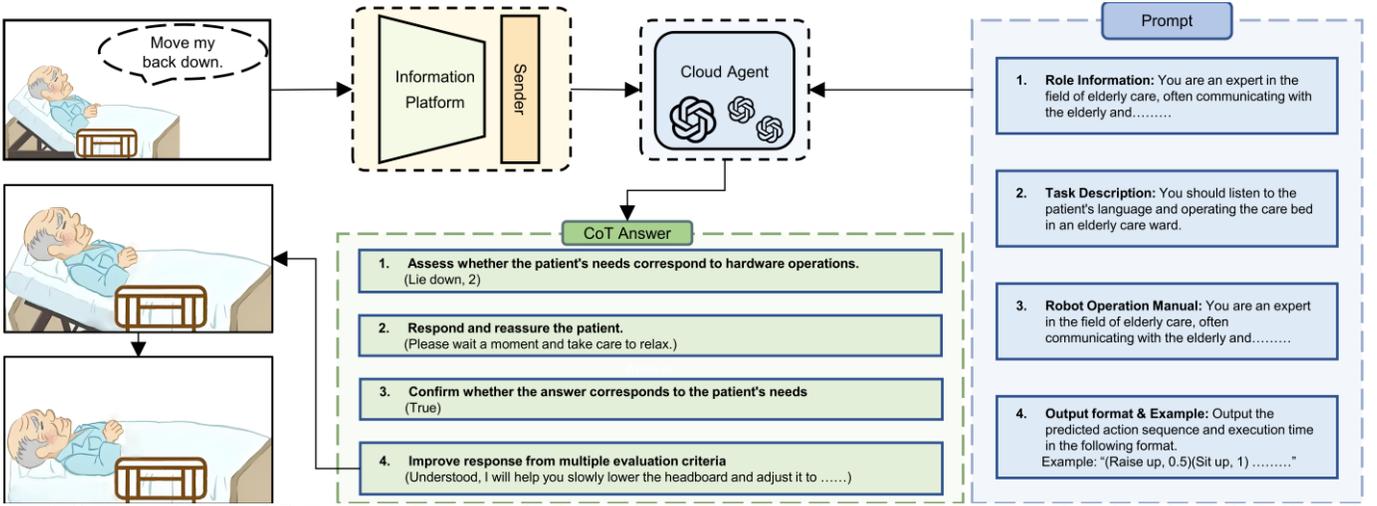

**Fig. 3.** Overview of the AoECR. The AoECR collects patient needs through the information platform and uploads the information to the EC Agent. EC Agent includes multiple LLMs, including a main LLM and several expert LLM. Through adversarial Chain of Thought (CoT) technology, the agent analyze patient needs to generate correct execution commands and humanized responses.

*B. Clarity-based data expansion*

Considering the specific needs of the geriatric care field, we expanded the dataset above to incorporate samples that reflect communication issues present in the elderly, including logical disorientation, inarticulateness, and stuttering. Specifically, we classified elderly's speech into four levels: high clarity, medium clarity, low clarity, and unclear. We defined the generated patient request sentences above as high clarity. Then we generated requests of medium clarity, low clarity, and unclear by obfuscating these sentences. For each high-clarity request, we prompted the LLM to generate different clarity levels, with the reserve of corresponding nurse responses. By the clarity-based expansion, we ultimately constructed a patient-nurse interaction dataset (PN-I) comprising 160,000 data pairs, which is suitable for the practical demands of elderly care.

## IV. Architecture of AoECR

This chapter will provide a detailed introduction to our proposed AoECR architecture, including the details of the EC Agent.

*A. Overall architecture*

The AoECR architecture we proposed primarily consists of two components: the information platform and the cloud agent. As shown in Fig. 3, the information platform is designed to collecting patient requests. These requests are subsequently uploaded to the cloud agent via an MQTT sender. In the cloud agent, we have deployed the EC Agent. The EC Agent is guided by a series of predefined prompts to describe its role information, task descriptions, and robot manuals. Additionally, a one-shot prompt is designed to clarify the output format and provide examples. Specifically, the EC Agent takes patient requests as input and responds step-by-step based on the designed chain of thought. With the help of the chain of thought, the EC Agent can generate control commands, confirm control commands, produce interactive responses, and perform optimization. After that, the control commands and interactive responses are sent back to the information platform and further distributed to the robots for execution and interaction.

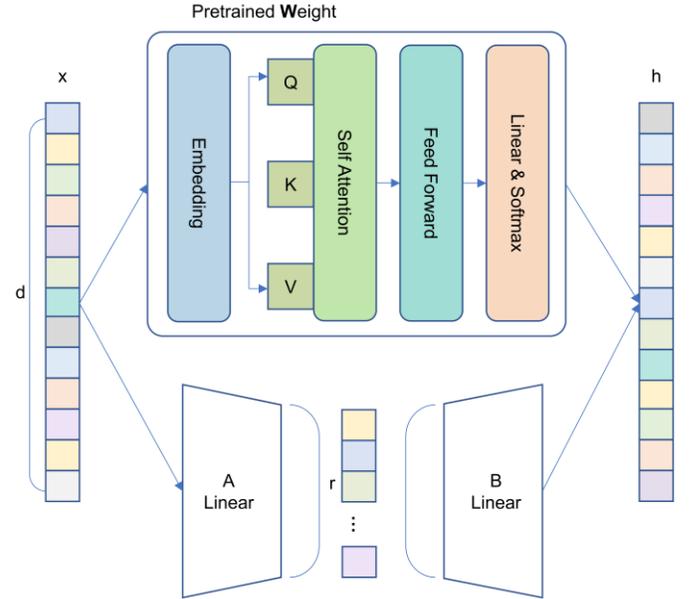

**Fig. 4.** Schematic of LoRA fine-tuning

*B. EC Agent*

The EC Agent comprises a fine-tuned main LLM, with a chain of self-check and an expert optimization. The main LLM is designed to generate tentative control commands and interactive responses, while the chain of self-check is designed to ensure the correct security of control commands, and the expert optimization is designed to optimize the interactive responses. In the main LLM, we selected the chatglm4-9b, developed by Tsinghua University, as our pre-trained model and employed the LoRA fine-tuning shown in Fig. 4. We trained two linear layers A and B, which are independent of the original model's pre-trained weights, to fine-tune the model's output.

The practical inference steps of the EC Agent are shown in Fig. 5. Initially, the main LLM process the patient's request to generate tentative control commands and tentative responses.

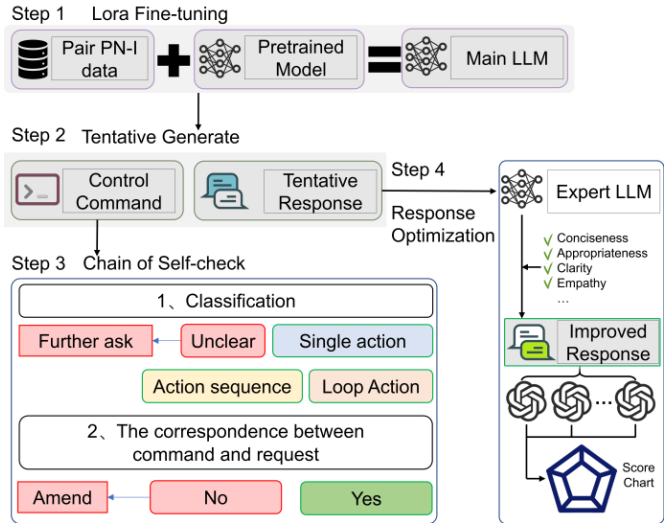

Fig. 5. The construction steps of the EC-Agent

Subsequently, we post-process the two outputs separately. The specific methods and steps for post-processing will be detailed in subsections C and D.

### C. Chain of self-check

We designed a CoS (Chain of Self-check) to ensure the accuracy and security of control commands. Firstly, we classified the patient's request into unclear, single action, action sequence, and loop action, which is necessary for control command generation. When the patient's request is unclear, we requested the main LLM to generate a further ask to obtain more specific request information from the patient. After that, the main LLM should assess the correspondence between the generated command and the patient's request. The CoS served as a safeguard to prevent mismatches between the command and the request. When found out a mismatch, the main LLM should revise its output.

### D. Expert optimization

To ensure the accuracy and clinical applicability of the responses, we designed an expert optimization for the post-processing of interactive responses, as depicted in Fig. 6. We deployed an expert LLM using appropriate prompts. Besides, we have designed 8 metrics for evaluating the nursing service, including conciseness, appropriateness, clarity, empathy, encouragement, explanation, safety, and understanding. After tentative responses are generated, we asked the expert LLM to optimize the tentative responses based on these 8 metrics to generate an optimized response. We designed an optimized equalizer here, to balanced different weights of the metrics. In the initial state, the weights of the metrics are balanced, and the expert LLM needs to balance each metric to optimize a compromised response. Furthermore, in actual clinical applications, we can request patients to score the responses of the agent. Based on the scores, we can get a human feedback equalizer. By adding the human feedback equalizer into the expert LLM, the optimization of response will reflect different personalized characteristics.

## V. EXPERIMENT

In this chapter, we will evaluate the proposed AoECR

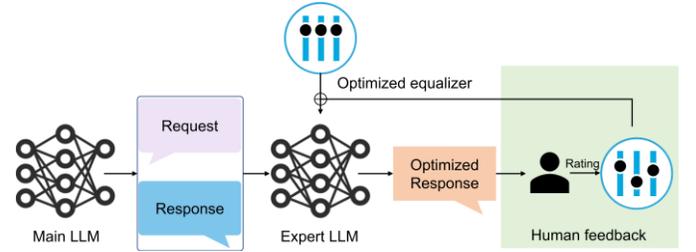

**Fig. 6.** Expert Optimization structure. The optimized equalizer is integrated into the Expert LLM through prompts.

architecture using quantitative assessment metrics and real-world application scenarios. The evaluation will include both control commands and interactive responses. To validate the effectiveness of the architecture in practical application settings, we will conduct demonstration through a series of physical examples.

### A. Elderly care nursing bed and AoECR deployment

An elderly care nursing bed is our control target, as shown in Fig. 7. The nursing bed is equipped with four actuating mechanisms and can perform eight distinct actions. Furthermore, we have established a communication link between the nursing bed and the information platform by employing the MQTT protocol. The information platform of AoECR is deployed on a personal computer, while the cloud agent is deployed on a server.

### B. Quantitative assessment

In the quantitative experiments, we conducted a separate quantitative evaluation for the control commands and interactive responses generated by the EC Agent.

Specifically, we utilized our PN-I dataset to test the correctness of control commands, with the pre-trained model chatglm4-9b serving as the baseline. Furthermore, we incrementally introduced prompt, fine-tuning, and CoS, and calculated the correctness of control commands at each stage.

In the process of quantifying the evaluation of interactive responses, we adopted a quantitative method similar with the expert optimization. Specifically, we constructed an evaluation environment composed of multiple expert LLMs. Each expert LLM was instructed to score each response based on the 8 metrics previously mentioned, with each criterion being scored on a scale of 1 to 5. Subsequently, we calculated the average of these scores as the quantitative score for each response. The detailed structure of this evaluation environment can be found in Fig. 5.

### C. Physical experiment

In the physical experiment, we deployed the AoECR on an elderly care nursing bed, with the experimenter playing the role of a patient lying on the bed. The patient expressed a series of requests through the information platform. After that, the platform uploaded the data to the EC Agent located on the server, which can process the requests, respond to the patient, and control the nursing bed to act. The experimental illustrations are presented in Fig. 8.

Furthermore, we observed that in actual scenarios, the degree of patient needs is different, which lead to different requirements for the execution time of actions. Consequently, we incorporated an additional prediction of action execution time into the CoS to achieve adaptation between action time and the degree of need. The corresponding experimental results are depicted in Fig. 9.

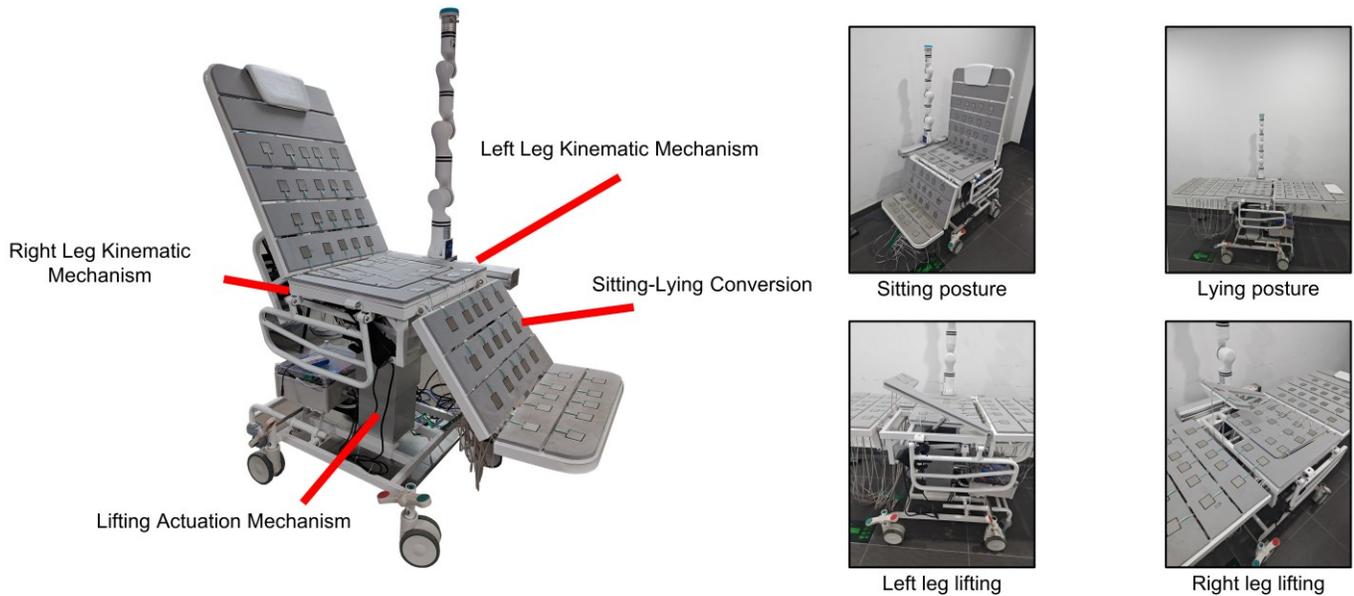

**Fig. 7.** Elderly Care Nursing Bed. Our nursing bed includes four executive agencies, including lifting actuation mechanism, sitting-lying conversion, left leg kinematic mechanism and right leg kinematic mechanism. The diagram on the right shows the limit states of different motion mechanisms.

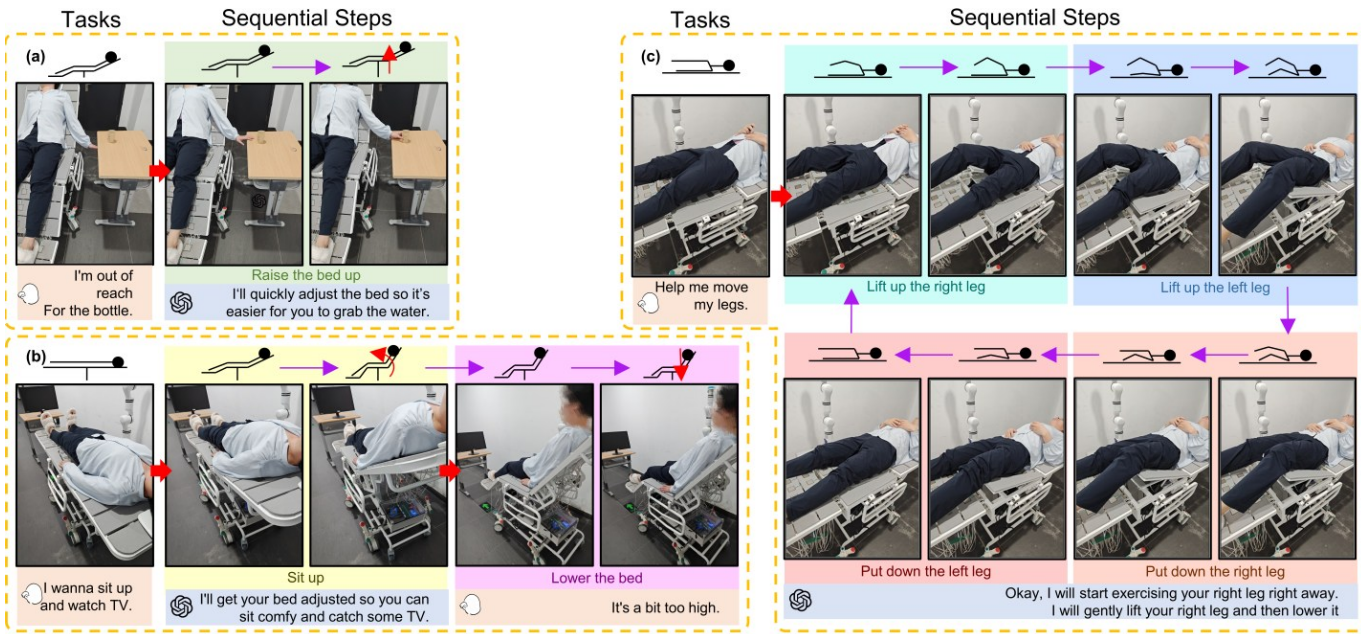

**Fig. 8.** Some examples of the action tasks. (i) The diagram delineates the actions that the agent performs with response. (ii) During task execution, patients can interrupt action at any time. (iii) Under special requirements, the agent can generate corresponding action sequences.

## VI. Discussion

In this chapter, we will conduct further discuss and analyze upon the experimental results, as well as outline potential directions for future research.

In the quantitative assessment experiments, we conducted a detailed quantification and comparison of the control commands and interactive responses generated by the EC Agent on the PN-I dataset. Specifically, to enhance the accuracy of control commands, we sequentially implemented Prompt, Fine-tune, and a CoS pipeline on top of the baseline model. We tested the performance of the EC Agent on the PN-I dataset and counted the accuracy in dialogue scenarios of varying clarity. The results are depicted in Fig. 10(a). Overall, Prompt achieves a 62.41% accuracy rate, initially meeting the requirements for hardware control. Fine-tune further increases the accuracy, and after confirmation by the Chain of Self-check pipeline, the total accuracy reaches 90.18%.

The accuracy statistics and comparison results for different clarity dialogue scenarios are shown in Fig. 10(b). It can be observed that after prompting and fine-tuning, the EC Agent achieves a 98.72% accuracy rate in high-clarity dialogue scenarios. However, the accuracy rate is low in medium, low, and unclear dialogue scenarios. The CoS pipeline, designed to address unclear dialogue issues, improves accuracy in all dialogue clarity scenarios. The CoS successfully identifies and corrects most erroneous tentative commands, achieving

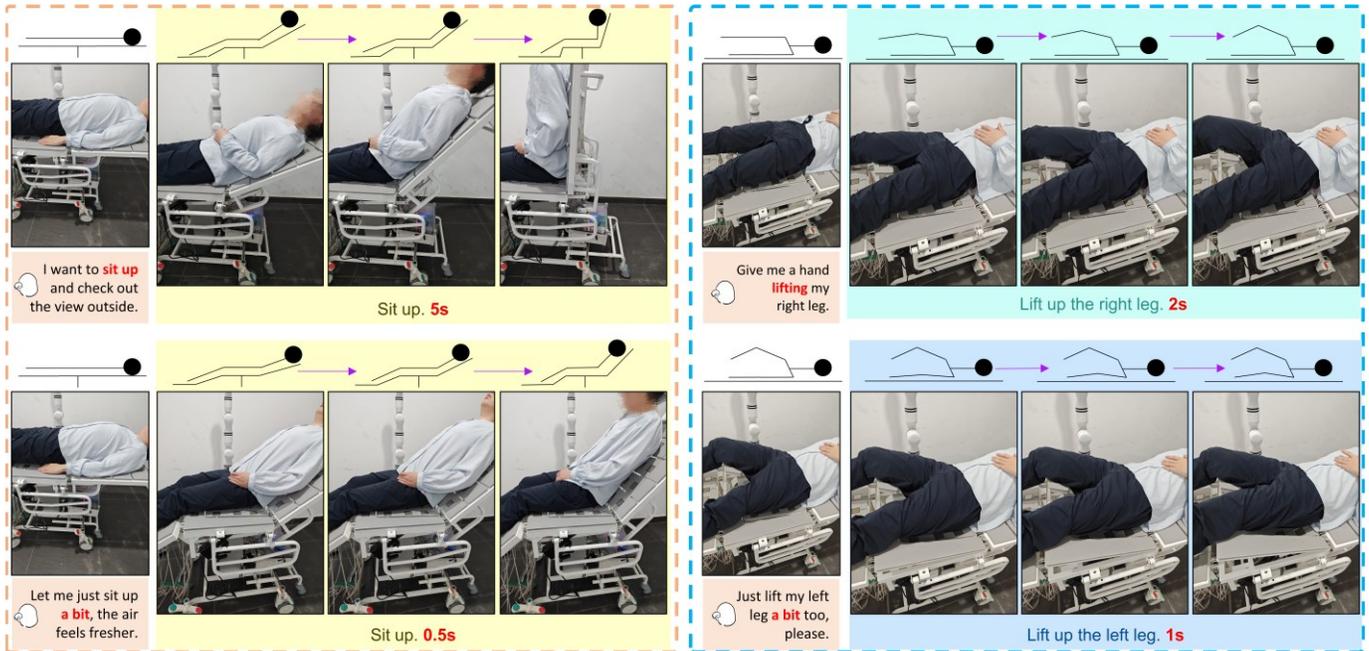

**Fig. 9.** Action time based on semantic differences. We present two sets of examples demonstrating that when patients' language conveys meanings akin to "slightly" or "a bit," there is a discernible variation in the execution time of actions.

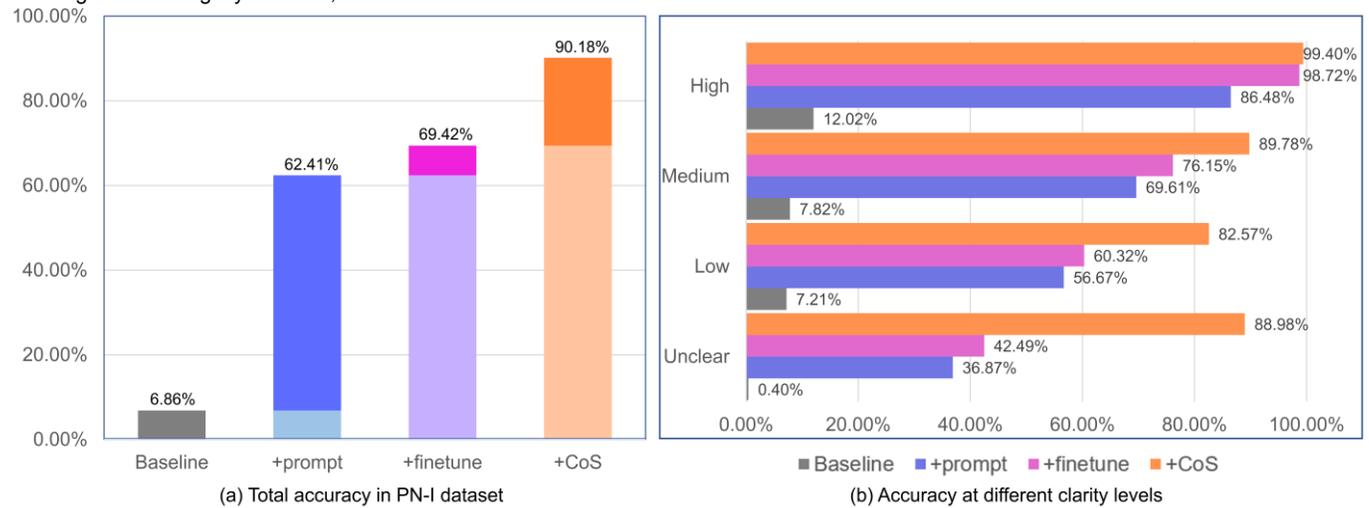

**Fig. 10.** Ablation experiment results: We count the accuracy of control commands. (a) shows the total accuracy, while (b) shows the accuracy at different clarity levels.

significant accuracy improvements in medium, low, and unclear clarity scenarios.

In the expert optimization phase, we employed an adjustable optimized equalizer and expert LLM to refine the tentative responses. The comparative experimental results in Fig. 11 validate that the optimized responses generated by expert optimization achieve significant improvements in terms of humanization and personalization.

Specifically, Fig. 11(a) presents the expert evaluation results for the tentative responses and the default optimized responses. Compared to the black tentative scores, the red scores after optimization show a significant improvement in most metrics. Due to the inclusion of more humanized information in the responses, there is a trade-off in conciseness. We have statistically represented the improvement percentage of scores for each metric on the histogram (depicted in blue), while the gray portion indicates the percentage of scores that remained unchanged. It demonstrates the effectiveness of expert optimization in the refinement of most responses.

In Fig. 11(b) and Fig. 11(c), we have selected different optimization equalizers within the expert optimization. Fig. 11(b) illustrates the effects of the optimization equalizer that prioritizes conciseness. Compared to the default optimization, the conciseness optimization achieves a significant enhancement in the conciseness metric while other metrics remain largely unchanged. The boxplot specifically displays the distribution of conciseness scores under tentative responses, default optimization, and conciseness optimization. The results indicate that the conciseness optimization maintains a level of conciseness like the tentative responses, preserving a high degree of brevity. Fig. 11(c) demonstrates the scores of a safety & encouragement optimization equalizer (indicated by the green section). The boxplot distributions of the safety scores and encouragement scores illustrate the improvements in these two metrics.

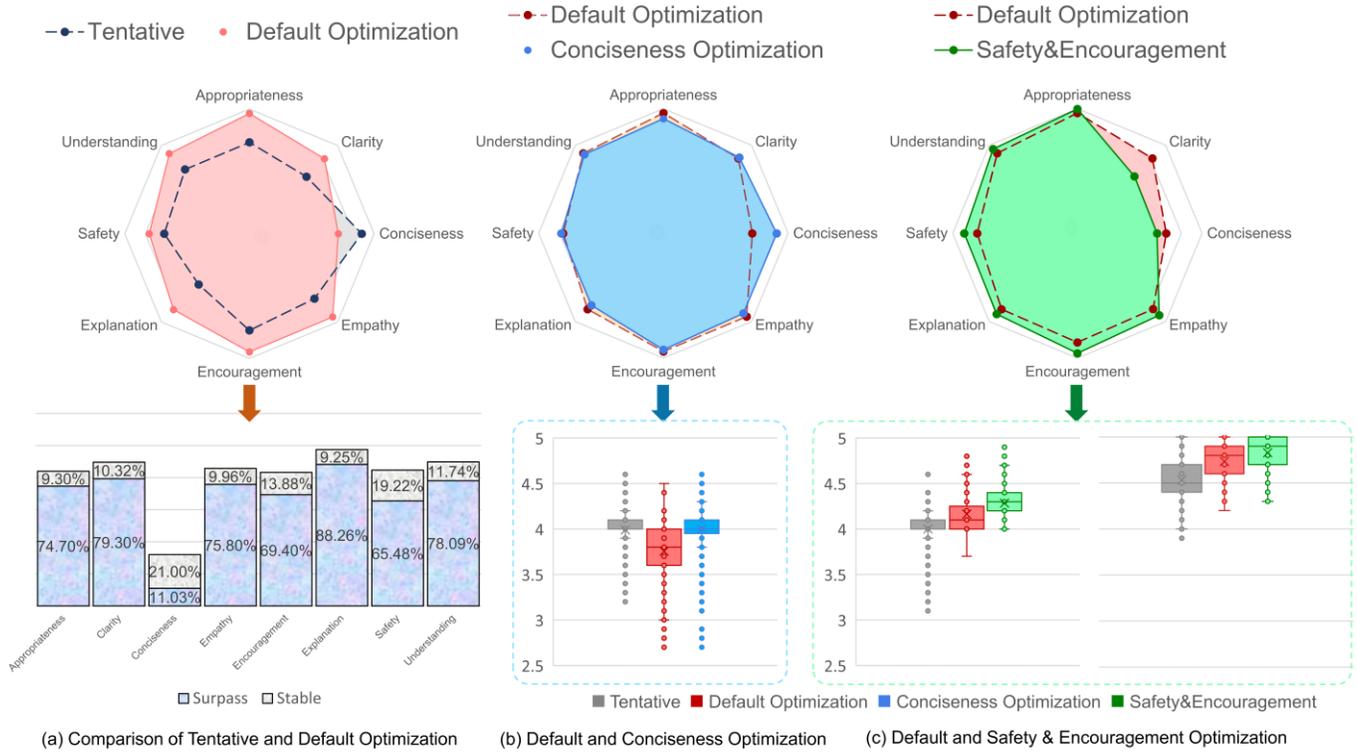

**Fig. 11.** Comparative results of Various Optimized Equalizers. We conducted three comparative experiments, which included comparisons between the tentative responses and the default equalizer (a), the default equalizer and the conciseness equalizer (b), as well as the default equalizer and the safety & encouragement equalizer (c).

In the physical experimentations, we deployed the AoECR architecture onto an elderly care nursing bed. The information platform serves as the input and output interface for the robot, while the EC Agent acts as the control center, akin to the brain of the system. The experimental illustrations are shown in Fig. 8 and Fig. 9. Concurrently, to provide a more comprehensive view of the experimental procedures, we have prepared a supplementary video. In the experiment, an experimenter played the role of a patient lying on the nursing bed and issued a series of requests verbally. Following the analysis and response generation by the AoECR, the nursing bed provided feedback to the patient and executed corresponding actions. Fig. 9 illustrates the execution results of the AoECR prediction of action time. Based on the varying degrees of demand expressed in the patient's language, the execution time of the nursing bed's actions also varied accordingly.

In summary, our AoECR architecture has successfully achieved the AI-ization of elderly care nursing beds. The quantitative experiments on control commands validated the effectiveness of various components within the EC Agent in improving the accuracy of control commands, demonstrating that the AoECR can generate correct control commands. Quantitative experiments on interactive responses verified their humanization. The distinct impact of different optimization equalizers on the metrics scores of responses indicates that the AoECR can produce personalized interactive responses. Physical experiments further confirmed that the elderly care robot, integrated with the AoECR, can respond to the patient's verbal requests with appropriate answers and actions. With the aid of the AoECR architecture, we have achieved direct interaction between the patient and the robot, successfully integrating artificial intelligence technology into the elderly care robot. It is noteworthy that our AoECR architecture is not limited to nursing beds. When the elderly care robot changes, we only need to modify the robot description in the LLM's prompt and create a dataset for fine-tuning to adapt the architecture to new robots.

However, the current work is limited to the semantic level, and the agent does not support multimodal input. In future work, we plan to replace the LLM in the EC Agent with a multimodal LLM. The expansion of the dataset is a critical step, while the information platform should be capable of supporting and processing various data formats. After proper multimodal fine-tuning, the new EC Agent will be able to support multimodal inputs, including visual and tactile information. With the aid of a multimodal agent, our AoECR can provide safer care for the elderly, achieving more comprehensive and complex nursing functions, and supporting a wider range of elderly care robots.

## VII. CONCLUSION

In this research, to make elderly care robots AI-ization, we developed a universal architecture for elderly care robots, which is called AoECR (AI-ization of Elderly Care Robot). Based on an elderly care nursing bed, our approach initially employed a zero-shot learning strategy to generate a patient-nurse interaction dataset. To make the dataset more suitable for the practical demands of elderly care, we extended the dataset based on the different clarity of expression. With the extension, the patient-nurse interaction dataset incorporates samples that reflect communication issues present in the elderly, including logical disorientation, inarticulateness, and stuttering. After that, to produce tentative control commands and responses, we designed an elderly care agent, comprising a main large language model fine-tuned on the dataset. Furthermore, we implemented a chain of self-check to ensure the accuracy and safety of control commands. Meanwhile, we designed an expert optimization module to refine tentative responses, which can enhance their humanization and personalization.

Our quantitative experiments showed an exceeding 90% accuracy in control commands and demonstrate the significate improvement of humanization and personalization in interactive responses, which can prove the efficacy of AoECR. We deployed the AoECR onto an elderly care nursing bed in the physical experiments, and confirmed its outstanding ability to facilitate patient-robot interaction.

The AoECR architecture can not only endow elderly care robots with AI capabilities but also facilitate direct interaction between patients and robots. With the help of the AoECR, we can potentially alleviate the workload on elderly care staff and enhance nursing efficiency in clinical settings. Moreover, the humanized interaction capabilities of the robots can reduce patient resistance, increase the practicality of elderly care robots, and foster their adoption and popularization. AoECR can be seen as a novel solution to the challenges associated with the AI-ization of elderly care robots. Our work lays the groundwork for more intelligent elderly care robots.

## VIII. Ethical Considerations

1. **Human subject ethics review approvals or exemptions:** Considering the elderly care robot under investigation in this study is currently in the prototype development phase and clinical testing has not yet been conducted, and there is no existing dataset that is fully applicable to this field, this paper proposes a method for constructing an zero-shot - generated dataset based on large - scale models to effectively address the issue of data scarcity. In summary, since the data utilized in this study do not originate from human subjects or case - related information, ethical approval is not required.
2. **Informed consent:** N/A.
3. **Privacy and confidentiality:** N/A.
4. **Compensation details:** N/A.